\documentclass[aps,twocolumn,showpacs,floatfix, bibnotes]{revtex4}

\usepackage{graphicx}%
\usepackage{dcolumn}
\usepackage{color}
\usepackage{amsmath}
\usepackage{here}

\begin{document}
\begin{sloppy}

\newcommand{\be}{\begin{equation}}
\newcommand{\ee}{\end{equation}}
\newcommand{\bea}{\begin{eqnarray}}
\newcommand{\eea}{\end{eqnarray}}
\newcommand\bibtranslation[1]{English translation: {#1}}
\newcommand\bibfollowup[1]{{#1}}

\newcommand\pictc[5]{\begin{figure}
                       \centerline{
                       \includegraphics[width=#1\columnwidth]{#3}}
                  \protect\caption{\protect\label{fig:#4} #5}
                    \end{figure}            }
\newcommand\pict[4][1.]{\pictc{#1}{!tb}{#2}{#3}{#4}}
\newcommand\rpict[1]{\ref{fig:#1}}

\newcommand\leqt[1]{\protect\label{eq:#1}}
\newcommand\reqtn[1]{\ref{eq:#1}}
\newcommand\reqt[1]{(\reqtn{#1})}

\newcounter{Fig}
\newcommand\pictFig[1]{\pagebreak \centerline{
                   \includegraphics[width=\columnwidth]{#1}}
                   \vspace*{2cm}
                   \centerline{Fig. \protect\addtocounter{Fig}{1}\theFig.}}

\title{Enhanced parametric processes in binary metamaterials}

\author{Maxim V. Gorkunov$^{1,2}$, Ilya V. Shadrivov$^1$, and Yuri S. Kivshar$^1$}

\affiliation{$^1$Nonlinear Physics Center, Research School of Physical Sciences and Engineering,
Australian National University, Canberra ACT 0200, Australia\\
$^2$Institute of Crystallography RAS, Moscow 119333, Russia}

\date{\today}

\begin{abstract}
We suggest double-resonant (binary) metamaterials composed of two
types of magnetic resonant elements, and demonstrate that in the
nonlinear regime such metamaterials provide unique possibilities
for  phase-matched parametric interaction and enhanced
second-harmonic generation.
\end{abstract}


\maketitle

Extensive studies of microwave properties of composite metallic
structures led to the experimental demonstration of left-handed
metamaterials~\cite{Shelby:2001-77:SCI} suggested long time
ago~\cite{Veselago:1967-2854:SPSS}. Such metamaterials are created
by resonant magnetic elements and operate for the wavelengths much larger than the period, allowing for the macroscopic effective
medium description. The parameters of the effective medium depend on the microscopic
structure of the metallic composites. Moreover, nonlinear response
of the metamaterial can become
substantial~\cite{Zharov:2003-37401:PRL, Lapine:2003-65601:PRE,
Lapine:2004-66601:PRE}, and their transmission characteristics can
be effectively controlled by external
fields~\cite{Gorkunov:2004-235109:PRB,
OBrien:2004-241101:PRB,Zharova:2005-1291:OE}.

Nonlinearities of metamaterials suggest their novel applications
such as frequency conversion~\cite{Lapine:2004-66601:PRE,
Shadrivov:physics/0506092}, tunable
transmission~\cite{Zharova:2005-1291:OE}, second-harmonic
imaging~\cite{Zharov:2005-91104:APL}, nonlinear beam focusing and
soliton propagation~\cite{Shadrivov:2005-S68:JOA}, etc. In contrast to
nonlinear optical media, composite metamaterials possess {\em nonlinear magnetic response} that can be engineered by inserting
nonlinear elements (e.g., diodes) into the resonant conductive
elements~\cite{Lapine:2003-65601:PRE, Zharov:2003-37401:PRL}.

In this Letter we suggest a novel type of composite metamaterials
with {\em double-resonant response} and demonstrate that in the nonlinear
regime such binary metamaterials are ideally suited for the first observation of the enhanced
phase-matched parametric interaction and second-harmonic
generation. Indeed, the quadratic nonlinear magnetic susceptibility
is proportional to a product of linear magnetic susceptibilities at
the frequencies of interacting waves. For conventional
single-resonant nonlinear metamaterials, the magnetic susceptibility
of the fundamental wave is relatively large, since it corresponds to
the backward wave near the
resonance~\cite{Shadrivov:physics/0506092} while the susceptibility
of the second-harmonic wave is rather small. In the metamaterial
with several resonances, it is possible to enhance the nonlinear
response, so that both linear susceptibilities of interacting waves
can become large.

To create a double-resonant metamaterial we suggest to mix two types
of resonant conductive elements (RCEs) with different resonant
frequencies~\cite{Chen:2004-5338:JAP}, as shown schematically in Fig.~\rpict{fig1} for the
structure consisting of two lattices of different split-ring
resonators. First, we study linear properties of the binary metamaterials. For
large wavelengths, each RCE can be described as a resonant circuit
(see, e.g., \cite{Gorkunov:2002-263:EPB,Marques:2002-144440:PRB})
characterized by self-inductance $L$, capacitance $C$, and
resistance $R$. We assume that the metamaterial consists of two
types of RCEs of the same shape (i.e., with the same $L$ and $R$),
but with different capacitances $C_1$ and $C_2$, and, thus,
different resonant frequencies.

\pict[0.65]{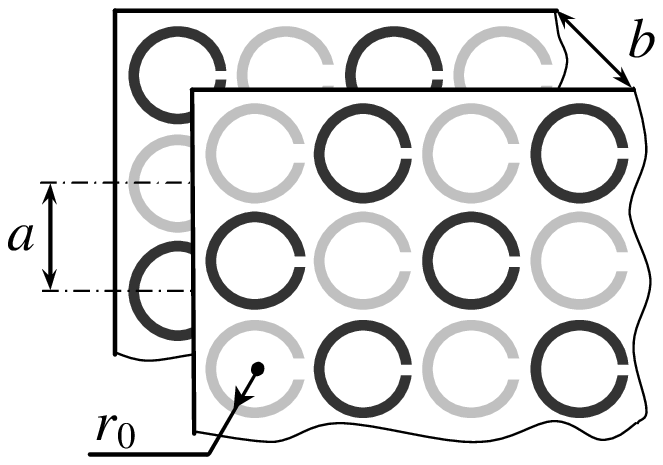}{fig1}{Schematic structure of binary metamaterials
with resonant magnetic elements of two types (black and gray).}

External homogeneous magnetic field $H_0$ applied perpendicular to
the RCE planes and oscillating with the frequency $\omega$ induces
the currents $I_1$ and $I_2$ in the resonators of the corresponding
type, which can be found from the impedance matrix
equation:
\bea\label{emf_short}
\mathcal E = Z_{1,2} ~I_{1,2} + \Xi~I_{2,1}
\eea
where $Z_{1,2}=Z^{(0)}_{1,2}-i \omega L_{11}$ and $\Xi=-i \omega
L_{12}$, $\mathcal E=i \mu_0 \omega S H_0$ is the electromotive force,
$S$ is the RCE area, $Z^{0}_\alpha \left( \omega \right) =-i \omega
L +i \left( \omega C_\alpha \right)^{-1} +R$ is the self-impedance
of an RCE of the type $\alpha$, $N_\alpha$ denotes the set of RCE
position indices of the type $\alpha$, and $M_{n^{\prime} n}$ are
the mutual inductances. The effective inductances are
\be \label{induct}
 L_{11}=\sum_{\substack{n, n^{\prime}\in N_1, \\ n^{\prime} \ne n}}
M_{n^{\prime} n} \ \text{;} \;\;
  \  L_{12}=\sum_{\substack{n \in N_1,\\ n^{\prime}\in N_2,}} M_{n^{\prime} n},
\ee

Solving the set of Eqs.~(\ref{emf_short}) with respect to the
currents, we obtain the magnetization of the metamaterial:
\be \label{M}
M = \frac{1}{2}~n S\left(I_1+I_2 \right) =
    n S^2 \mu_0 \mathcal {K} \  H_0,
\ee
where $\mathcal K = i\omega \left(Z_1 + Z_2 - 2 \Xi \right) / 2
\left( Z_1 Z_2 - \Xi^2 \right)$, and  $n=(a^2 b)^{-1}$ is the total
volume density of RCEs. Using the general relation for magnetic
induction of media in the external field, $B=\mu_0 \left( H_0+2/3
M\right)$ (see Ref.~\cite{Gorkunov:2002-263:EPB} for details) and
definition of the magnetic susceptibility $\chi$, $M=\chi H$, we
calculate the magnetic permeability $\mu$,
\be\label{chi}
\mu\left( \omega\right) = 1+\chi = 1+\dfrac{\mu_0 n
S^2}{ \mathcal {K}^{-1} - \mu_0 n S^2/3}.
\ee
In the case $C_1=C_2$, the result (\ref{chi}) reduces to that
obtained previously for single-resonant
structures~\cite{Gorkunov:2002-263:EPB}.

In Fig.~\rpict{fig2}(a), we plot the permeability vs. frequency for
typical parameters: RCE radius $r_0=2 \text{mm}$, wire thickness
$l=0.1~\text{mm}$, which gives self-inductance $L=8.36~\text{nHn}$
(see \cite{Gorkunov:2002-263:EPB}). To obtain RCEs of the type 1
with the resonant frequency of $ \omega_{01}=6 \pi \cdot10^9
\text{rad/s}$ ($\nu_0=3 \text{GHz}$), we take $C_1=0.34~\text{pF}$.
The resonance frequency of the type 2 RCEs is chosen as $\omega_{02}
= X \omega_{01}$ with $X=1.75$, i.e., $C_2=C_1/X^2$. The lattice
constants are $a=2.1 r_0$ and $b=0.5 r_0$. The RCE quality factor,
$Q=\omega_{01} L/R$, can reach the values up to
$10^3$~\cite{Shelby:2001-77:SCI}. However, by inserting diodes this
value may decrease, and therefore we take $Q=300$.

\pict{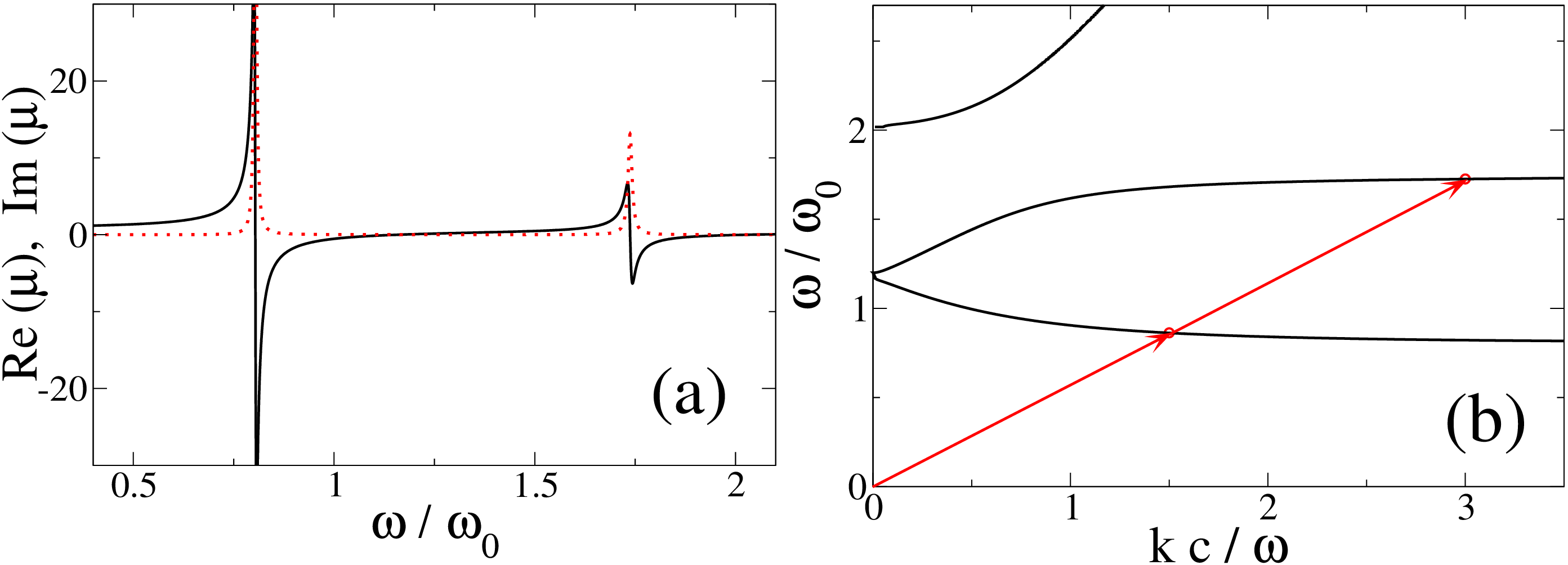}{fig2}{(a) Real (solid) and imaginary (dashed)
parts of magnetic permeability of the binary metamaterial. (b)
Spectrum of electromagnetic waves. Arrows show the perfectly
phase-matched second-harmonic generation.}

Figure~\rpict{fig2}(a) confirms that indeed in such structures there
exist two resonances and two frequency ranges with negative magnetic
permeability. Positions of the macroscopic resonances are shifted
from the resonant frequencies of individual RCEs; the shift is not
the same for two resonances, and the resulting ratio of the resonant
frequencies is about 2.17.

Nonlinear metamaterials can be created by inserting nonlinear
elements. In order to obtain a material with low resistive losses,
it is preferable using variable capacitance insertions, varactor
diodes~\cite{Lapine:2003-65601:PRE}. We assume that the capacitance
of RCEs (both linear and nonlinear) is determined by varactors, and
the difference between two types of resonators arises due to
different varactors.

A general expression for the voltage drop on a varactor can be
written in the form~\cite{Lapine:2004-66601:PRE,
Gorkunov:2004-235109:PRB}
\be\label{Uins}
U(t) = I(t) R_{\text{ins}} \left[ U(t) \right] +
\frac{1}{C_{\text{ins}}\left[ U(t) \right]}
\int\limits_{-\infty}^{t} I(t')\, d t',
\ee
and it can be simplified in the limit of low current and low
voltage. We assume that resistance is constant, while capacitance
can be expanded into Taylor series, $C_\text{ins}\left(U \right)
\simeq C \left(1+\gamma U \right) $. The corresponding
solution of Eq.~(\ref{Uins}) gives a linear capacitive contribution
to the impedance as well as weak quadratic nonlinearity.

For the three-wave parametric processes, we write $I_\alpha \left(t
\right) =\sum_{\nu=1}^{3} I_\alpha \left(\omega_\nu \right)\exp{\left(
-i \omega_\nu t\right)}+c. c.$, $\omega_1=\omega_2+\omega_3$, and
the nonlinear analogue of Eq.~(\ref{emf_short}) takes the form:
\begin{multline}\label{emf_nonl}
\mathcal E\left(\omega_{1} \right) = Z_{1,2}\left(\omega_{1,2} \right)
    ~I_{1,2}\left(\omega_{1,2} \right)+
    \Xi\left(\omega_{1,2} \right)~I_{2,1}\left(\omega_{1,2} \right)+\\
    \frac{\gamma_{1,2}}{C_{1,2}^2 \omega_2 \omega_3}
    I_{1,2}\left(\omega_2 \right)I_{1,2}\left(\omega_3 \right).
\end{multline}
Since the nonlinear part of the capacitance is much smaller than the
linear one, we apply an iterative procedure and use linear currents calculating nonlinear contribution to obtain
\begin{multline}\label{chi2}
\chi^{\left(2 \right)}\left(\omega_1 ; \omega_2, \omega_3\right) = \dfrac{4 \ i
\ \chi\left(\omega_1 \right) \chi\left(\omega_2 \right) \chi\left(\omega_3
\right) }{\mu_0 S^3 n^2 \omega_1 \omega_2 \omega_3} \times \\
\sum_{\alpha} \frac{\gamma_{\alpha}}{C_{\alpha}^2}
A_{\alpha} \left(\omega_1 \right)
A_{\alpha} \left(\omega_2 \right)
A_{\alpha} \left(\omega_3 \right),
\end{multline}
where
\begin{eqnarray} \label{A}
A_{1,2} \left(\omega \right) =
    \dfrac{Z_{2,1} \left(\omega \right) -
    \Xi\left(\omega \right)}{ Z_1 \left(\omega \right) +
    Z_2 \left(\omega \right)-2\Xi \left(\omega \right)},
\end{eqnarray}
characterize the relative contribution from the currents,
$I_1\left(\omega \right)$ and $I_2\left(\omega \right)$, to the
total magnetization of the metamaterial, and $A_1+A_2=1$. In the
limit of identical varactors in both sublattices, i.e., $C_1=C_2$
and $\gamma_1=\gamma_2$, Eq. (\ref{chi2}) coincides with the results
obtained in Ref.~\cite{Lapine:2003-65601:PRE}.

We note that zeros of the denominator in Eq.~(\ref{A}) are canceled
out by zeros of linear susceptibility in numerator and they do not
lead to any increase of the nonlinear response. At the same time,
the resonant poles of the linear magnetic susceptibilities in
Eq.~(\ref{chi2}) lead to a dramatic increase of nonlinear
interaction when the wave frequencies approach resonances. Clearly,
the effect is stronger when all the waves are close to magnetic
resonances of the metamaterial. Therefore, the binary metamaterial
provides an unique possibility for realizing this general concept,
as we show below for the example of SHG.

As has been shown recently~\cite{Shadrivov:physics/0506092}, SHG in
the media with negative refraction differs from the conventional
scheme. In particular, it is possible to satisfy the phase-matching
conditions for counter-propagating waves. As a result, a
semi-infinite sample of a nonlinear quadratic metamaterial operates
as a frequency converting mirror reflecting radiation with the
double frequency of the incident wave. Remarkably, in the lossless
regime the conversion efficiency is close to 100\%. In a more
realistic case of a finite-size metamaterial slab, high efficiency
is possible for the slabs of several tens of wavelengths.

For the double-resonant medium, first we analyze the spectrum of
electromagnetic waves, $\omega(k)$. We consider the waves with the
magnetic field perpendicular to the planes of resonators and assume
that the electric component of the metamaterial generates a
plasma-like dielectric response, $\varepsilon\left(\omega \right) =
1 - \omega_\text{p}^2 / \omega^2$, where the plasma frequency
$\omega_\text{p}=1.2 \omega_0$ is selected between two magnetic
resonances. The wave spectrum has three branches, as shown in
Fig.~\rpict{fig2}(b). Two branches, which are close to the
 magnetic resonances, correspond to large wavenumbers. Importantly, we
can find the points of the exact phase-matching between fundamental
and second-harmonic waves, for both waves close to the resonances.

We consider the case of normal incident wave propagating along
$z$-axis, and present the magnetic field inside the sample using
slowly varying amplitudes:
\begin{multline}\label{H}
H\left(z, t \right)=a_1^+(z) e^{-i k z-i \omega t} + a_2^+ (z) e^{-2 i k z-2 i
\omega t} + \\
a_1^-(z) e^{ i k z-i \omega t} + a_2^- (z) e^{2 i k z-2 i \omega t}
+ c.c.,
\end{multline}
where $k=\omega/c \sqrt{\varepsilon(\omega) \mu^{\prime}(\omega)}$
(as usual $\mu=\mu^\prime+ i \mu^{\prime \prime}$), the phase
mismatch $\Delta=\varepsilon(\omega)
\mu^{\prime}(\omega)-\varepsilon(2\omega) \mu^{\prime}(2\omega)$ is
assumed to be small. The coupled-mode equations for the amplitudes
$a_1^+$ and $a_2^+$  are written in the form
\bea\label{couple1}
\frac{d a_1^+}{d \zeta} + \kappa_1 a_1^+ = i \sigma_1 a_2^+ {a_1^+}^*, \nonumber\\
\frac{d a_2^+}{d \zeta} + \kappa_2 a_2^+ = -i \sigma_2 {a_1^+}^2
{a_1^+}^*, \eea
where we use the notations $\kappa_1 = \mu^{\prime \prime}(\omega)
{\varepsilon (\omega)}^{1/2}/ 2 {\mu^{\prime }(\omega)}^{1/2}$,
$\kappa_2 = \left[ i \Delta -  \varepsilon(2 \omega)
\mu^{\prime \prime}(2 \omega) \right] \left[ \varepsilon (\omega)
\mu^{\prime }(\omega) \right]^{-1/2}$, $\sigma_1 = 0.5\left[
\varepsilon (\omega) / \mu^{\prime }(\omega) \right]^{1/2} \chi^{(2)}
(\omega; 2\omega, -\omega)$, $\sigma_2 = \varepsilon (2 \omega)
\left[\mu^{\prime }(\omega) \varepsilon (\omega) \right]^{-1/2}
\chi^{(2)} (2\omega; \omega, \omega)$,
and $\zeta=\omega z /c$ is the dimensionless coordinate. Equations for
the amplitudes $a_1^-$ and $a_2^-$ are the same as equations for
$a_1^+$ and $a_2^+$, except the opposite signs of the spatial
derivatives. We solve these equations numerically with appropriate
boundary conditions and obtain the dependence of the SH reflection
coefficient, i.e., the ratio of the reflected energy flux of the SH
to the incident wave, as a function of the ratio of the two resonant
frequencies $X$, shown in Fig.~\rpict{refl_new}(a) for three slab
thicknesses. Calculating results shown in Fig.~\rpict{refl_new}(a),
we were adjusting the frequency of the incident wave to satisfy the
phase-matching conditions. Large $X$ correspond to non-resonant
limit, when the SH field is not in resonance. Decreasing $X$ we
drive both FF and SH waves closer to the magnetic resonances, and
the conversion rate increases. At the same time, losses become
stronger, and finally they dominate suppressing SHG efficiency. For
small relative shifts (below $X=1.75$), the phase matching cannot be
archived. The incident field amplitude and nonlinear coefficients
$\alpha_1 = \alpha_2$, were chosen in such a way that maximum
nonlinear modulation in simulations was $\chi^{(2)} (\omega;
2\omega, -\omega) H_\omega < 0.2$. Such modulation is expected in resonant nonlinear processes, since even in realistic non-resonant case~\cite{Lapine:2004-66601:PRE}, the nonlinear modulation of $0.01$ was created by the external magnetic fields with amplitudes less then 1 A/m. Our results demonstrate that for a one-wavelength-thick slab, the SHG enhancement due to the second resonance can become larger by at least one order of magnitude. The decrease of losses would allow increasing the efficiency.

Dependence of the maximum reflection coefficient of the SH wave and
reflection coefficient in non-resonant case ($X=3$) on the slab
thickness is shown in Fig.~\rpict{refl_new}(b). One can see that the
major relative increase of the SHG process in resonance, compared to
non-resonant case, is observed for thin nonlinear slabs.

\pict{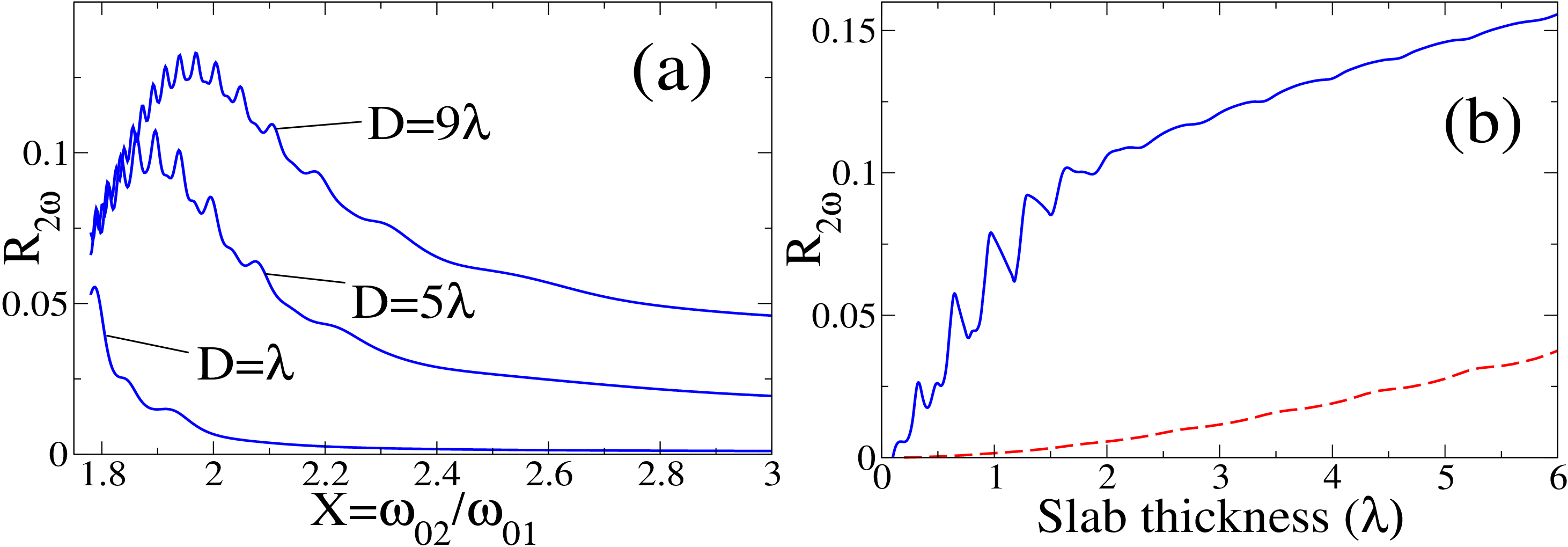}{refl_new}{(a) Reflection coefficient of the second
harmonics as function of resonant frequency ratio $X$, for different
slab thicknesses $D$. (b) Maximum reflection coefficient of the
second harmonics (solid) and reflection coefficient at $X=3$
(dashed) as function of slab thickness.}

In conclusion, we have suggested double-resonant metamaterials for the study of phase-matched parametric interactions in composite
nonlinear media. In particular, we have analyzed a composite
structure consisting of two types of resonant magnetic elements, and
demonstrated that such a binary resonant structure can enhance
significantly the second-harmonic generation.

\end{sloppy}
\end{document}